\documentclass[prd, aps, nofootinbib, preprint, letterpaper]{revtex4-2}

\usepackage{mathtools}
\usepackage{color}
\definecolor{darkred}{RGB}{196,0,0}

\usepackage{amsmath, amssymb, amsfonts}

\usepackage[hidelinks]{hyperref}
\usepackage{ulem}
\usepackage{setspace}
    
\newcommand{\be}{\begin{equation}}
\newcommand{\ee}{\end{equation}}
\newcommand{\ba}{\begin{eqnarray}}
\newcommand{\ea}{\end{eqnarray}}

\newcommand{\uh}{\hat \nu}

\newcommand{\oh}{\hat\omega}

\begin{document}

\title {Energy loss of a heavy fermion in a collisional QED plasma}

\author{Yun Guo$^*$}
\author{Luhua Qiu}
\author{Ruizhe Zhao}
\affiliation{Department of Physics, Guangxi Normal University, Guilin, 541004, China}
\affiliation{Guangxi Key Laboratory of Nuclear Physics and Technology, Guilin, 541004, China}

\author{Michael Strickland$^\dagger$}
\affiliation{Department of Physics, Kent State University, Kent, Ohio 44242, United States}

\renewcommand{\thefootnote}{\fnsymbol{footnote}}
\footnotetext[1]{yunguo@mailbox.gxnu.edu.cn}
\footnotetext[2]{mstrick6@kent.edu}
\renewcommand{\thefootnote}{\arabic{footnote}}

\begin{abstract}
We compute the energy loss of heavy fermions moving in a plasma, taking into account the modification of the photon collective modes induced by collisions using a Bhatnagar-Gross–Krook collisional kernel.  We include contributions from both hard and soft scatterings of the heavy fermion using a collisionally modified hard-thermal-loop resummed propagator.  Using this method, one does not need to introduce a separation scale between hard- and soft-momentum exchanges.  To place our calculation in context, we review other theoretical approaches to computing the collisional energy loss of fermions and discuss the systematics and results obtained in each approach compared to using a resummed propagator for both hard and soft momentum exchanges.  Our final results indicate that self-consistently including the effect of collisions in the self-energies of the resummed propagator results in an increased energy loss compared to using collisionless hard-thermal-loop propagators.  The effect becomes larger as the magnitude of the coupling constant and the velocity of the fermion increase.
\end{abstract}

\maketitle

\newpage

\section{Introduction}

The study of energy loss in high-temperature plasmas is important for understanding the jet suppression in the quark-gluon plasma (QGP) generated in relativistic heavy-ion collisions \cite{Cunqueiro:2021wls,Apolinario:2022vzg}.  In his seminal work on this topic, Bjorken predicted that collisional energy loss would result in suppression of jets in relativistic heavy-ion collisions, which began the consideration of this as a key signature for the production of a QGP~\cite{Bjorken:1982tu}.  It was later realized that radiative parton energy loss would be the dominant mechanism for the suppression of jets at asymptotically high collision energies \cite{Gyulassy:1990ye,Gyulassy:1990bh,Wang:1991hta,Wang:1992qdg,Gyulassy:1993hr,Baier:1996kr,Zakharov:1996fv,Gyulassy:2000fs,Wiedemann:2000za,Wang:2001ifa,Arnold:2001ms,Arnold:2002ja,Arnold:2002zm,Gyulassy:2003mc,Djordjevic:2003zk,Djordjevic:2006tw,Majumder:2010qh,Bellac:2011kqa,Mehtar-Tani:2013pia,Blaizot:2015lma,Qin:2015srf}; however, at currently achievable heavy-ion collision energies at the Relativistic Heavy Ion Collider (RHIC) and the Large Hadron Collider (LHC), both collisional and radiative energy loss are important and must be included to properly interpret experimental observations of the suppression of jets in such collisions \cite{Qin:2007rn,Schenke:2009ik,Schenke:2009gb}. In particular, when considering heavy quarks, due to the dead-cone effect, elastic scatterings that induce collisional energy loss are more important to take into account~\cite{Dokshitzer:2001zm,Zhang:2003wk,Mustafa:2004dr,Wicks:2005gt}. When both effects are included, one can understand the observations of jet suppression at both RHIC and LHC collision energies in a manner consistent with expectations from perturbative quantum chromodynamics (QCD)~\cite{PHENIX:2001hpc,STAR:2002ggv,STAR:2002svs,STAR:2009ojv,PHENIX:ppg090,ATLAS:2010isq,CMS:2011iwn,ATLAS:2012tjt,ALICE:2015mjv}.

Early works on collisional energy loss of high-energy partons propagating through the QGP included hard or soft momentum exchanges separately within perturbative QCD~\cite{Bjorken:1982tu,Gyulassy:1990bh,Thoma:1990fm,Mrowczynski:1991da}.  In Ref.~\cite{Braaten:1991we}, Braaten and Thoma presented a systematic method for including both hard and soft momentum exchanges by introducing a momentum separation scale $q^\star$, above which a diagrammatic calculation of heavy-quark energy loss with bare propagators was performed, and below which a hard-thermal-loop (HTL)  \cite{Braaten:1989mz,Braaten:1989kk,Braaten:1991gm} resummed propagator was used.  The inclusion of this separation scale made the calculations in the hard and soft sectors manifestly finite in the infrared (IR) and ultraviolet (UV) limits, respectively; however, the hard part was logarithmically IR divergent as $q^\star$ decreased and the soft part was logarithmically UV divergent as it increased. Braaten and Thoma showed that these two divergences canceled exactly, so that the sum of the hard and soft parts was independent of the separation scale at leading order in the QCD coupling constant.  It was also found that without introducing the separation scale, the divergent contributions in the hard and soft parts also canceled when the dimensional regularization was used~\cite{Carignano:2021mrn,Comadran:2023vsr}.

Despite this key progress, an issue remained with the calculation of Braaten and Thoma, namely that for small velocities that were still within the region of applicability of their calculation, their asymptotic evaluation of the integrals lead to the collisional energy loss being negative in an equilibrium QGP, resulting in energy gain instead of energy loss at small velocities.  This unphysical behavior was later eliminated by direct numerical evaluation of the necessary integrals that appear at leading order in the coupling constant \cite{Romatschke:2003vc,Romatschke:2004au}.  One caveat of the method introduced in Refs.~\cite{Romatschke:2003vc,Romatschke:2004au} was that at finite gauge coupling, there was a residual dependence on the cutoff scale $q^\star$ separating hard and soft momentum exchanges.  The residual dependence went to zero as the gauge coupling constant went to zero.  At large values of the coupling constant relevant to QGP physics, the dependence of the total collisional energy loss on the separation scale $q^\star$ allowed the authors to quantify the theoretical uncertainty associated with the introduction of a theta function like separation scale $q^\star$ between the hard and soft contribution to the collisional energy loss.

An alternative approach to computing collisional energy loss was proposed in the original paper of Braaten and Thoma~\cite{Braaten:1991we}, which consisted of evaluating it using the diagrammatic method, but using HTL-resummed propagators in the $t$-channel diagrams for all momentum exchanges.  This alternative did not require the introduction of a separation scale and would give a manifestly IR and UV finite result; however, it was not implemented in their paper.  The method was eventually applied by Djordjevic and Gyulassy in Ref.~\cite{Djordjevic:2006tw}, allowing them to compute the collisional energy loss without having to resort to separate calculations in the hard and soft sectors.  Similarly to Refs.~\cite{Romatschke:2003vc,Romatschke:2004au} they found that this approach eliminated the unphysical energy gain at low velocities (momentum). 

In all of these prior works \cite{Braaten:1991we,Romatschke:2003vc,Romatschke:2004au,Djordjevic:2006tw} the authors made use of HTL-resummed propagators that were obtained in the collisionless limit from computations of the HTL self-energies.  However, when collisions are included, the HTL self-energies are modified, resulting in direct damping of the quasiparticle modes and a shift of the Landau damping cut into the lower half of the complex energy plane \cite{Carrington:2003je,Schenke:2006xu,Zhao:2023mrz}.  In Refs.~\cite{Carrington:2003je,Schenke:2006xu,Zhao:2023mrz}, the effect of collisions on the soft-scale self-energies was performed using a number-conserving Bhatnagar-Gross–Krook (BGK) collisional kernel.  This collisional kernel is a modified form of the relaxation time approximation collisional kernel and models collisions through the inclusion of a collision rate $\nu$.  Using the same method as used by Djordjevic and Gyulassy \cite{Djordjevic:2006tw}, in this paper we compute the collisional energy loss using the diagrammatic method, but using BGK-modified HTL (BGK-HTL) self-energies instead of collisionless HTL self-energies.  In this way, we can self-consistently include the effect of collisions on soft- and hard-momentum exchanges without introducing an explicit separation scale. 

This is to be contrasted to prior work using BGK-HTL self-energies.  In Refs.~\cite{Han:2017nfz,Shi:2018aeb}, the authors computed only the soft contribution to collisional energy loss of heavy quarks using the BGK-HTL self-energies.  In Refs.~\cite{YousufJamal:2019pen,Jamal:2020emj}, the authors also computed only the soft contribution to collisional energy loss of heavy quarks within a quasiparticle model of QCD at zero and finite chemical potential.   Finally, we mention that Ref.~\cite{Elias:2014hua} considered the effect of a finite relaxation time on the soft contribution to collisional energy loss using a polarization tensor that was derived within an effective hydrodynamic theory.  In all of these previous works, the authors found that the inclusion of collisional effects into the resummed gauge propagator resulted in an increase in the collisional energy loss.  

Our work goes beyond these prior studies by including the hard contribution to the collisional energy loss in a self-consistent manner.  In addition to this, we emphasize again that the formalism we use does not require the introduction of an explicit separation scale for hard and soft momentum exchanges and we do not approximate the integrals using asymptotic limits.  As a consequence, similarly to Refs.~\cite{Romatschke:2003vc,Romatschke:2004au,Djordjevic:2006tw} we avoid the problem of unphysical energy gain. Finally, to assess the dependence of our results on the calculational scheme used, we provide explicit comparisons between the collisional energy loss obtained using the Braaten-Thoma \cite{Braaten:1991we} and Romatschke-Strickland \cite{Romatschke:2003vc,Romatschke:2004au} methods.  In order to demonstrate the general method, we focus herein on the calculation in QED since this is somewhat more straightforward than the full QCD calculation and postpone the consideration of the full QCD calculation to a forthcoming paper.  Within this context, we prove that the resulting energy loss is gauge-independent and evaluate it numerically.  Our final results indicate that, in QED using couplings consistent with those expected to be generated in the QGP ($\alpha_s \sim 0.3$), the inclusion of collisional effects in the gauge boson propagator results in an approximately 10\% increase in the heavy fermion collisional energy loss at high momentum.

The structure of our paper is as follows. In Sec.~\ref{re}, we make systematic comparisons between the theoretical methods for computing the collisional energy as developed in Refs.~\cite{Braaten:1991we,Romatschke:2003vc,Romatschke:2004au,Djordjevic:2006tw} and demonstrate that different methods lead to the same result in the weak-coupling limit, while a moderate discrepancy exists for a realistic QCD coupling constant relevant at temperatures not far above the critical temperature, where the effect of collisions among medium partons is expected to be more pronounced. In Sec.~\ref{elwithcolli}, we carry out the calculation of the collisional energy loss of a heavy fermion propagating through a hot QED plasma by using a resummed gauge-boson propagator that uses the BGK-HTL self-energies. With a phenomenological estimate of the collision rate entering into the BGK collisional kernel, we present our numerical results for the energy loss with emphasis on the enhancement caused by the collision effect. In addition, we compare our results with results obtained in prior works. Finally, our conclusions and outlook are presented in Sec.~\ref{con}.

\section{Theoretical methods to compute the collisional energy loss of a heavy fermion in a hot plasma}\label{re}

Considering a high-energy fermion with mass $M$ and momentum ${\bf p}$ propagating through a hot QED plasma at a temperature $T$, it may lose energy through interactions with the medium partons. The rate of energy loss $d E/d x$ per distance traveled is given by
\be
-\frac{d E}{d x}=\frac{1}{v}\int_M^\infty
dE^\prime (E-E') \frac{d \Gamma}{d E'}\, ,
\ee
where the velocity of the incident heavy fermion with energy $E$ is given  by ${\bf v}={\bf p}/E$ and  the interaction rate $\Gamma(E)$ can be expressed in terms of the Feynman diagrams. For example, the contribution to $\Gamma(E)$ from scattering by a thermal electron is given by
\ba
\Gamma (E)&=&\frac{1}{2E} \int \frac{d^3 {\bf{p}}'}{(2\pi)^3 2{E}'}\int \frac{d^3 {\bf k}}{(2\pi)^3 2k} n_{F}(k)\int \frac{d^3 {\bf {k}}'}{(2\pi)^3 2{k}'}[1 - n_{F}({k}')]\,\nonumber \\
&& \hspace{4cm} \times (2\pi)^4\delta^4(P+K-{P}'-{K}')\,\bigg(\frac{1}{2}\sum_{\mathrm{spins}}|\mathcal{M}|^2\bigg)\, .
\label{interaction_rate}
\ea
In the above equation, $P=(E, {\bf p})$ and $P^\prime=(E^\prime, {\bf p}^\prime)$ are the four-momenta of the incoming and outgoing fermion, respectively. The four-momenta of the medium partons that scatter off the incident fermion are denoted by $K=(k, {\bf k})$ and $K^\prime=(k^\prime, {\bf k}^\prime)$. In addition, the phase space is weighted by a Fermi-Dirac distribution $n_{F}(k)=(e^{k/T}-1)^{-1}$ and a Pauli-blocking factor $1- n_{F}(k^\prime)$ for the incoming and outgoing electrons, respectively. Similarly, to get the contribution to $\Gamma(E)$ from scattering by a thermal photon, one should use the Bose-Einstein distribution $n_{B}(k)=(e^{k/T}+1)^{-1}$ and replace $1- n_{F}(k^\prime)$ with the Bose-enhanced factor $1+ n_{B}(k^\prime)$ in the above equation. To obtain the energy loss $-d E/d x$, one only needs to insert $(E-E')/v\equiv \omega/v$ into the integrand of the above equation for the interaction rate.

\begin{figure}[htbp]
\begin{center}
\includegraphics[width=0.8\linewidth]{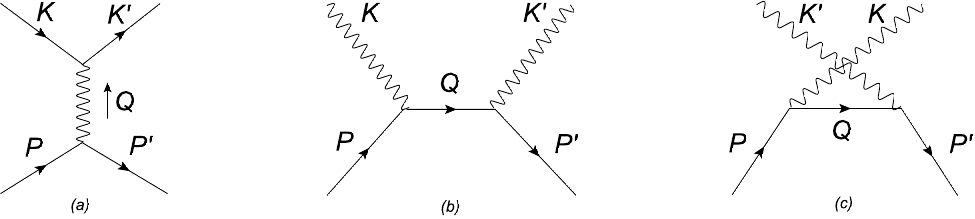}
\caption{Leading order Feynman diagrams for the elastic scattering $e^- \mu\rightarrow e^- \mu$ and $\gamma \mu\rightarrow \gamma \mu$. (a) $t$-channel Coulomb scattering. (b) $s$-channel Compton scattering. (c) $u$-channel Compton scattering.}
\label{lod}
\end{center}
\end{figure}

Let us focus on elastic scattering $e^-\mu\rightarrow e^-\mu$ where the incident fermion is assumed to be a massive muon. 
For the hard scattering process with large momentum transfer $\sim T$, we can only consider the tree-level Feynman diagram, as shown in Fig.~\ref{lod}. On the other hand, when the momentum of the exchanged photon is on the order of $e T$, self-energy insertion into the bare photon propagator has to be taken into account. Thus, for the soft process, one needs to use an effective photon propagator, i.e., the hard thermal loop resummed propagator. In covariant gauge, it reads 
\be\label{resumpro}
D^{\mu\nu}(Q)=\frac{1}{Q^2-\Pi_T(\oh)} A^{\mu\nu}+\frac{1}{q^2-\Pi_L(\oh)} \frac{\omega^2 q^2}{Q^4}B^{\mu\nu}-\frac{\eta}{Q^4}Q^{\mu} Q^{\nu}, \,  
\ee
where $\eta$ is the gauge parameter and the four momentum of the exchanged photon is denoted by $Q=K^\prime-K=(\omega, {\bf q})$. The transverse and longitudinal part of the photon self-energy are given by
\be
\Pi_L(\oh) = m_\gamma^2 \bigg(-1+\frac{\oh}{2}\ln\frac{\oh+1+i \epsilon}{\oh-1+i \epsilon}\bigg)\,,\quad \Pi_T(\oh) = \frac{\oh^2}{2}m_\gamma^2\bigg(1-\frac{\oh^2-1}{2\oh}\ln\frac{\oh+1+i \epsilon}{\oh-1+i \epsilon}\bigg)
\, .
\label{se}
\ee
which are complex valued for ${\hat \omega}^2<1$ and the two projectors are defined as
\be
A^{\mu\nu}=-g^{\mu\nu}+\frac{Q^{\mu} Q^{\nu}}{Q^2}+\frac{{\tilde M}^{\mu} {\tilde M}^{\nu}}{{\tilde M}^2}\,,\quad\quad\quad B^{\mu\nu}=-\frac{Q^2}{(M\cdot Q)^2}\frac{{\tilde M}^{\mu} {\tilde M}^{\nu}}{{\tilde M}^2}\,.
\ee
In the above equations, $\oh\equiv \omega/q$ and the screening mass is defined by $m_\gamma^2=e^2 T^2/3$. In addition, $M^{\mu}$ is the heat bath vector, which in the local rest frame is given by  \mbox{$M^{\mu}=(1,0,0,0)$}. The part that is orthogonal to $Q^\mu$ is denoted as 
\be
{\tilde M}^\mu=M^\mu-\frac{M\cdot Q}{Q^2} Q^\mu\, .
\ee

At large momentum transfer, the resummed propagator is reduced to the bare one provided that the coupling constant is small. As a result, the squared matrix element computed based on the resummed propagator is expected to be valid for both hard and soft processes. Thus, in terms of the integral variables $k$, $q$, and $\omega$, the collisional energy loss of the fast fermion can be expressed as
\ba\label{elnew}
-\bigg(\frac{dE}{dx}\bigg)&=&\frac{e^4}{4\pi^3 v^2}\int_{0}^{\infty} d k\,  n_F(k) \Big(\int_{0}^{2k/(1+v)} dq\,q^2 \int_{-v q}^{v q} d\omega \, \omega+\int_{2k/(1+v)}^{2k/(1-v)} dq\,q^2 \int_{q-2k}^{v q} d\omega \, \omega\Big) \nonumber\\&& \hspace{1.5cm} \times \bigg[ |\Delta_L(Q)|^2 f_1(k,q,\omega)+\frac{Q^4}{q^4} |\Delta_T(Q)|^2  (f_2(k,q,\omega)-f_1(k,q,\omega))\bigg]\, ,
\ea
where
\be\label{protl}
\Delta_T(Q)=\frac{1}{Q^2-\Pi_T(\oh)}\,,\quad\quad\quad \Delta_L(Q)=\frac{1}{q^2-\Pi_L(\oh)}\, ,
\ee
and
\ba
f_1(k,q,\omega)&=& 2 k\frac{\omega+k}{q^2}+\frac{\omega^2/q^2-1}{2}\,,\nonumber \\
f_2(k,q,\omega)&=& 3\frac{k^2+k\omega+\omega^2/4}{q^2}-(1-v^2)\frac{k^2+k \omega +q^2/2}{q^2-\omega^2}-\frac{v^2}{4}\, .
\ea
To obtain the above result, we make use of the fact that in an isotropic medium, the energy loss is independent of the direction of ${\bf v}$. In addition, besides the assumption $M\gg T$, we also assume $v\gg T/E$ and $E\ll M^2/T$. These assumptions, together with the energy and momentum conservation lead to a constraint on the energy of the outgoing medium parton, $k^\prime \sim T$. As a result, the transferred energy $\omega=k^\prime-k$ and momentum ${\bf q}={\bf k^\prime}-{\bf k}$ are also on the order of $T$ or even smaller. To leading order in $T/M$, we can take $P\approx P^\prime$ and the contributions to the energy loss from Compton scattering\footnote{The Compton scattering in QED involves the $s$- and $u$-channel diagrams as shown in Fig.~\ref{lod}. Notice that the muon is not thermalized due to its large mass and one can use the bare fermion propagator to compute the matrix element for these two channels.} are suppressed by $(T/M)^2$ which have been neglected in our calculation. On the other hand, when the transferred energy and momentum is very large, for example, $q \sim E$, one should consider the opposite limit $E\gg M^2/T$ which corresponds to the ultrarelativistic limit $v\rightarrow 1$. In this case, a complete treatment of the collisional energy loss can be found in Ref.~\cite{Peigne:2007sd} where the Compton scattering cannot be neglected anymore.

The above method which uses a resummed gluon propagator in the calculations of the squared matrix element has been adopted in Ref.~\cite{Braaten:1991we} as an alternative way to study the soft contributions to $-d E/d x$. Then it has been generalized to arbitrary momentum exchange in Ref.~\cite{Djordjevic:2006tw}. In addition to a well-defined energy loss, the most important advantage of this method is that there is no need to introduce an artificial cutoff $q^\star$ to define the so-called hard and soft contributions to the collisional energy loss. 

\begin{figure}[htbp]
\begin{center}
\includegraphics[width=0.49\linewidth]{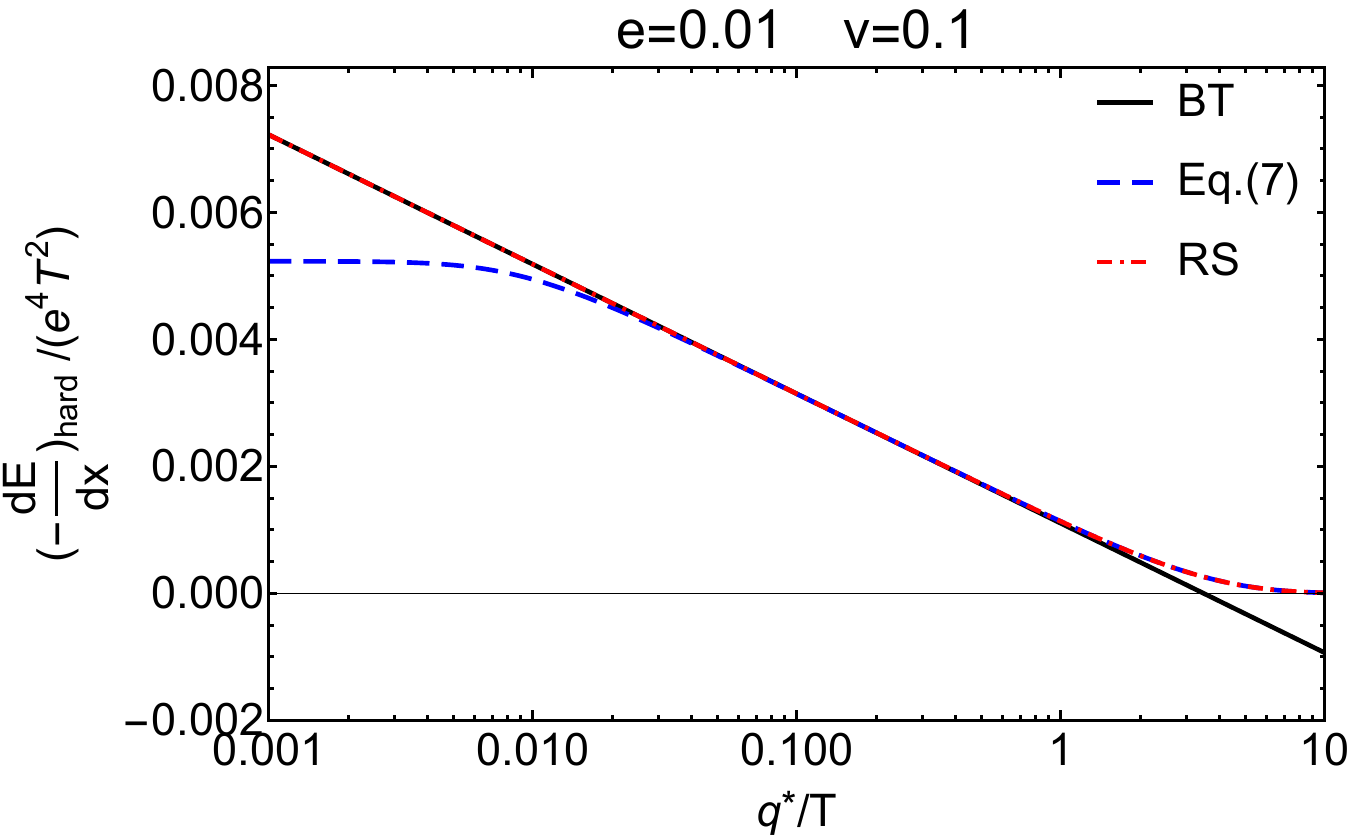}
\includegraphics[width=0.49\linewidth]{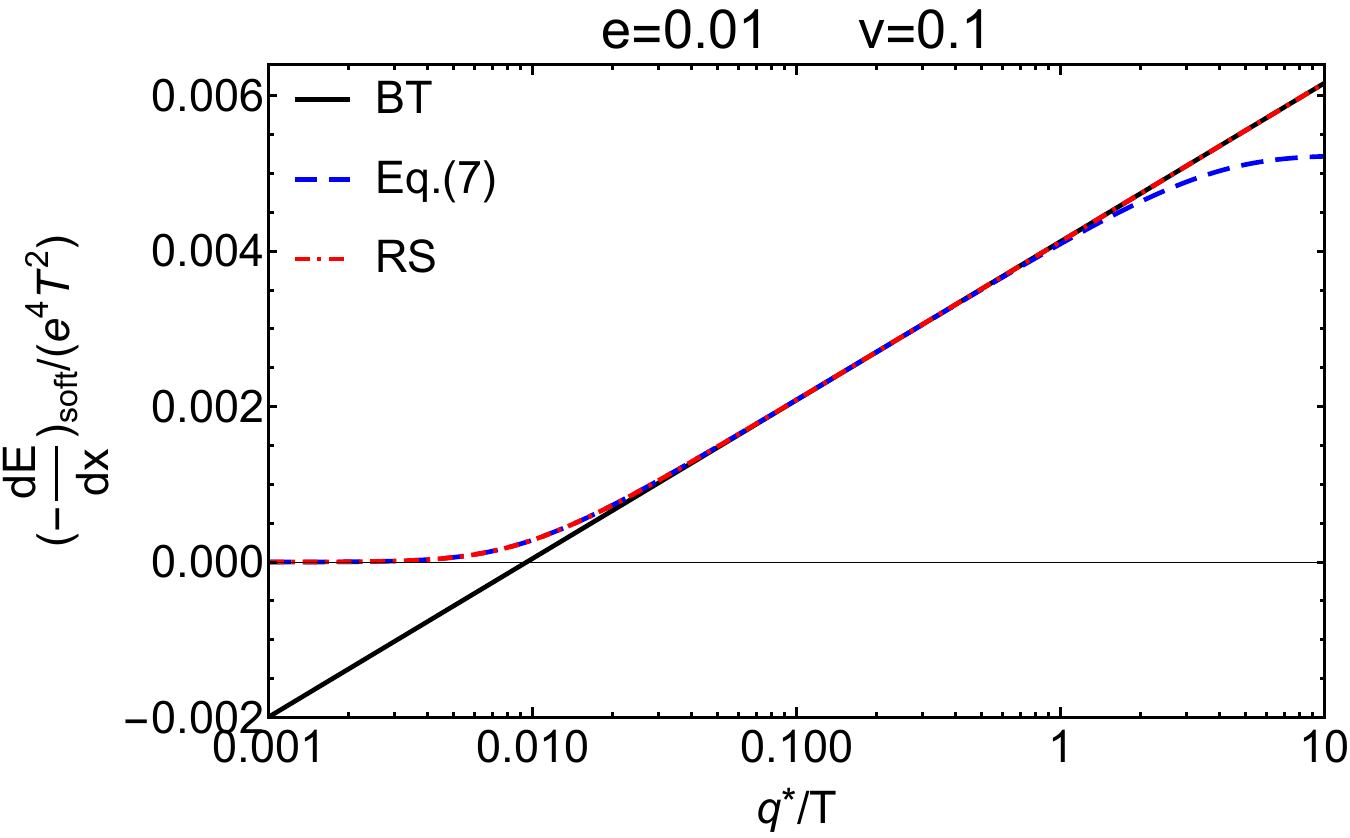}
\caption{Hard and soft contributions to the collisional energy loss as a function of $q^\star/T$ obtained from different methods.}
\label{comqs}
\end{center}
\end{figure}

In Refs.~\cite{Braaten:1991we,Romatschke:2003vc}, by following a similar procedure as the above, the hard contributions to $-d E/d x$ are obtained with the resummed propagator replaced by the bare one. The result is given by
\ba
-{\Big(\frac{dE}{dx}\Big)}_{\rm {hard}} &=&\frac{e^4}{4\pi^3 v^2}\int_{0}^{\infty} d k\,  n_F(k) \Big(\int_{0}^{2k/(1+v)} dq\,q^2 \int_{-v q}^{v q} d\omega \, \omega+\int_{2k/(1+v)}^{2k/(1-v)} dq\,q^2 \int_{q-2k}^{v q} d\omega \, \omega\Big) \nonumber\\&\times&\frac{2\omega}{(\omega^2-q^2)^2}\bigg[2(k-{\bf v}\cdot {\bf k})^2+\frac{1-v^2}{2}(\omega^2-q^2)\bigg]\delta(\omega-{\bf v}\cdot {\bf q})\, .
\label{hard}
\ea
However, an infrared divergence would appear as $q\rightarrow 0$, and thus an extra constraint $\theta(q-q^\star)$ on the integral variable $q$ has to be introduced. Explicitly, we need to perform the following integrals
\ba\label{changeva}
\frac{1}{2(2\pi)^2}\int \frac{d^3 \mathbf{k}}{k} \int \frac{d^3\mathbf{k}'}{k^\prime} \theta(q-q^\star) &\to& \int_{\frac{1+v}{2}{q^\star}}^{\infty} d k\int_{ q^\star}^{\frac{2k}{1+v}} q dq \int_{-v q}^{v q} d\omega 
+\int_{\frac{1+v}{2}{ q^\star}}^{\infty} d k\int_{\frac{2k}{1+v}}^{\frac{2k}{1-v}}q dq \int_{q-2k}^{v q} d\omega 
\nonumber \\&+&\int_{\frac{1-v}{2}{q^\star}}^{\frac{1+v}{2}{q^\star}} d k\int_{q^\star}^{\frac{2k}{1-v}} q dq \int_{q-2k}^{v q} d\omega\,.
\ea
In Ref.~\cite{Braaten:1991we}, by assuming $q^\star/T\ll 1$, the following integrals haven been used instead
\be\label{changevav2}
\frac{1}{2(2\pi)^2}\int \frac{d^3 \mathbf{k}}{k} \int \frac{d^3\mathbf{k}'}{k^\prime} \theta(q-q^\star) \to\int_{0}^{\infty} d k\int_{ q^\star}^{\frac{2k}{1+v}} q dq \int_{-v q}^{v q} d\omega 
+\int_{0}^{\infty} d k\int_{\frac{2k}{1+v}}^{\frac{2k}{1-v}} q dq \int_{q-2k}^{v q} d\omega \, .
\ee
Clearly, those $q^\star$'s in (\ref{changeva}) which do not lead to divergence have been set to be zero in (\ref{changevav2}). This is valid due to the fact that the typical momentum of the medium partons is on the order of $T$. Therefore, in the small $q^\star$ region, the hard contribution to the energy loss from Ref.~\cite{Braaten:1991we} which used (\ref{changevav2}) to perform the integrals (denoted as BT result) agrees with that from Ref.~\cite{Romatschke:2003vc} which used (\ref{changeva}) to perform the integrals (denoted as RS result). However, as the cutoff gets smaller, both results show a logarithmic enhancement, see Fig.~\ref{comqs} and have an obvious discrepancy as compared the hard contribution based on Eq.~(\ref{elnew})\footnote{To make comparisons among different theoretical methods, we introduce $\theta(q-q^\star)$ in Eq.~(\ref{elnew}) to define the corresponding hard contribution. Similarly, with $\theta(q^\star-q)$, one can obtain the soft contribution based on Eq.~(\ref{elnew}). Notice that in both cases, the squared matrix element is obtained using the resummed HTL propagator.}. This is actually very easy to understand, since for a soft momentum exchange, it is necessary to use the resummed propagator, which regulates the infrared divergence. In fact, in the limit $q^\star\rightarrow 0$, the result based on Eq.~(\ref{elnew}) corresponds to the total energy loss. On the other hand, the hard contribution is expected to vanish as $q^\star \rightarrow \infty$. In this limit, 
BT result becomes negative as the assumption $q^\star/T\ll 1$ does not hold any more.

\begin{figure}[htbp]
\begin{center}
\includegraphics[width=0.7\linewidth]{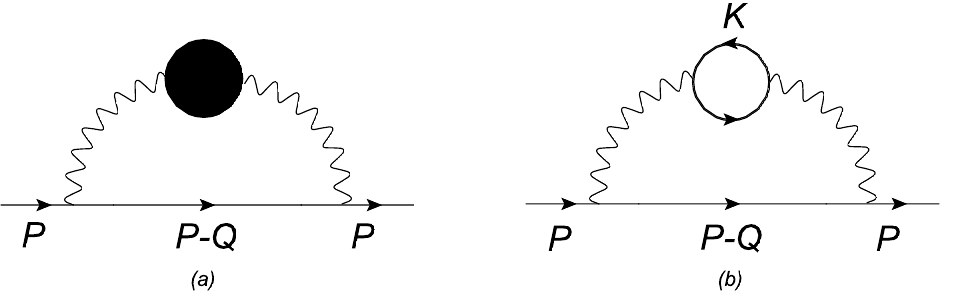}
\caption{Feynman diagram for the fermion self-energy. (a) with resummed photon propagator. (b) two-loop diagram.}
\label{mse}
\end{center}
\end{figure}

The interaction rate $\Gamma$ can be also expressed as
\ba\label{inrav2}
\Gamma(E)&=&-\frac{1}{2E} {\rm Tr}\, \big[({\displaystyle{\not}}P+M) {\rm Im }\,\Sigma (P)\big]\, \nonumber \\
 &=&- \frac{ e^2}{4 \pi^2 v} \int_0^{\infty} dq\, q  \int_{-v q}^{v q} d \omega (1+n_B(\omega))\, {\rm Im} \Big[ \Delta_L(Q)+(v^2-\oh^2)\Delta_T(Q)\Big]\, ,
\ea
where the fermion self-energy $\Sigma(P)$ is given by the Feynman diagram in Fig.~\ref{mse}(a). In Ref.~\cite{Braaten:1991we}, by using the HTL resummed photon propagator, the soft contribution to $-d E/d x$ is found to be
\be
-{\Big(\frac{dE}{dx}\Big)}_{\rm {soft}}=\frac{ e^2 m_\gamma^2}{8\pi v^2} \int_0^{q^\star} dq  \int_{-v q}^{v q} d \omega \omega^2 \Big[|\Delta_L(Q)|^2+\frac{1-\oh^2}{2}(v^2-\oh^2)|\Delta_T(Q)|^2\Big] \, .
\label{soft}
\ee
In the above equation, we have expanded the Bose-Einstein distribution function $n_B(\omega)$ for $\omega\sim e T \ll T$ and kept only the nonvanishing leading order contribution. It can be proven that Eq.~(\ref{soft}) is equivalent to the soft contributions obtained in Ref.~\cite{Romatschke:2003vc}, where the energy loss was calculated based on the classical energy loss formula~\cite{Thoma:1990fm}. On the other hand, by introducing a cutoff $q^\star$ for the transferred momentum $q$, the integrals in Eq.~(\ref{elnew}) become 
\ba\label{changevasoft}
\frac{1}{2(2\pi)^2}\int \frac{d^3 \mathbf{k}}{k} \int \frac{d^3\mathbf{k}'}{k^\prime} \theta(q^\star-q) &\to& \int_{\frac{1+v}{2}{q^\star}}^{\infty} d k\int_{0}^{ q^\star} q dq \int_{-v q}^{v q} d\omega + \int_0^{\frac{1+v}{2}{q^\star}} d k\int_{0}^{\frac{2k}{1+v}} q dq \int_{-v q}^{v q} d\omega 
\nonumber \\&&\hspace{-1cm} +\int_0^{\frac{1+v}{2}{ q^\star}} d k\int_{\frac{2k}{1+v}}^{\frac{2k}{1-v}}q dq \int_{q-2k}^{v q} d\omega 
+\int_{\frac{1-v}{2}{q^\star}}^{\frac{1+v}{2}{q^\star}} d k\int^{q^\star}_{\frac{2k}{1+v}} q dq \int_{q-2k}^{v q} d\omega\,. \nonumber \\
\ea
For soft processes, the upper limit of $q$ should be much smaller than $T$. Therefore, only the first term contributes and the above integrals can be simplified as
\be\label{changevasoft2}
\frac{1}{2(2\pi)^2}\int \frac{d^3 \mathbf{k}}{k} \int \frac{d^3\mathbf{k}'}{k^\prime} \theta(q^\star-q) \to \int_{0}^{\infty} d k\int_{0}^{ q^\star} q dq \int_{-v q}^{v q} d\omega\,.
\ee
Notice that although $|\mathcal{M}|^2$ becomes complicated with the use of the resummed propagator, Eq.~(\ref{elnew}) can be further simplified by requiring that the integrand should be symmetric in $\omega$. Consequently, the integral over $k$ can be carried out analytically, which gives the prefactor $m_\gamma^2$ in Eq.~(\ref{soft}). In fact, it can be easily shown that with ($\ref{changevasoft2}$), the energy loss from Eq.~(\ref{elnew}) is identical to Eq.~(\ref{soft}).

At this point, it is also useful to extend the above analysis to the case of QCD. The corresponding soft contributions which come from the $t$-channel quark-quark and quark-gluon scatterings can be obtained from Eq.~(\ref{soft}) by replacing $e$ by the strong coupling $g$, multiplying by a color factor $C_F=4/3$, and replacing the screening mass $m_\gamma^2$ with its QCD counterpart $m_D^2=g^2 T^2 (1+N_f/6)$, where $N_f$ is the number of light flavors in the medium~\cite{Braaten:1991we}. The same result should be obtained when the interaction rate $\Gamma(E)$ is defined in terms of the squared matrix element as shown in Eq.~(\ref{interaction_rate}), which is then computed by using the resummed gluon propagator. According to the origin of the prefactor $m_\gamma^2$ in Eq.~(\ref{soft}), we can expect the following change in QCD 
\be\label{masschange}
2 e^2 \int k n_F(k) d k \rightarrow 2
g^2 n_f \int   \frac{2}{3}  k\, n_F(k) d k+ g^2 \int 4 k \, n_B(k) d k\, .
\ee
It can be easily shown that the above integrals lead to the screening mass $m_\gamma^2$ for QED and $C_F m_D^2$ for QCD, up to a same and trivial constant. In Eq.~(\ref{masschange}), the color factors for the quark-quark and quark-gluon scatterings are given by $2/3$ and $4$, respectively. In the case of a Fermi-Dirac distribution function, there is an extra factor of $2$ because scatterings from thermal positrons or antiquarks also need to be included. Based on the above discussions, the squared matrix element computed with resummed gluon propagator becomes identical for the quark-quark and quark-gluon scatterings provided that the momentum of the exchanged gluon is soft.

If one further assumes $q^\star \gg e T$, the integrand in Eq.~(\ref{soft}) can been expanded. Keeping the leading order results, the $q^\star$-dependent part in the soft contributions can be analytically calculated. An important conclusion in Ref.~\cite{Braaten:1991we} is that the cutoff dependence is completely canceled between the hard and soft contribution\footnote{Based on Eq.~(\ref{soft}), the expanded energy loss is referred to as the BT result in Ref.~\cite{Braaten:1991we}, while the unexpanded one is the RS result in Ref.~\cite{Romatschke:2003vc}.}. However, the expansion breaks down when $q^\star \sim e T$, so the BT result becomes negative as $q^\star \rightarrow 0$ where the energy loss approaches zero due to the increasingly smaller integral region in Eq.~(\ref{soft}). See Fig.~\ref{comqs} for a numerical demonstration.

For very large $q^\star$, on the other hand, Fig.~\ref{comqs} also shows that the soft contribution from Eq.~(\ref{soft}) has a logarithmic divergence, while the result from Eq.~(\ref{elnew}) is finite and corresponds to the total energy loss as $q^\star \rightarrow \infty$. It is not surprising to see such an unphysical behavior because Eq.~(\ref{soft}) only holds in the HTL approximation and thus is not valid for a hard process. In contrast, the interaction rate given in Eq.~(\ref{inrav2}) is general provided that one uses a resummed photon propagator without the HTL approximation. Taking into account the hard contribution to $-d E/d x$, Eq.~(\ref{inrav2}) can be expanded because the self-energy correction to the (inverse) bare-photon propagator can be treated as a small perturbation. To leading order in the expansion, $\Gamma(E)$ corresponds to the Feynman diagram as shown in Fig.~\ref{mse}(b) which relates to the imaginary part of the one-loop photon self-energy ${\rm Im}\, \Pi(K)$. Clearly, the calculation of the self-energy has to be carried out beyond the HTL approximation, and the exact result for ${\rm Im}\, \Pi(K)$ involves the Fermi-Dirac distribution functions $\sim (n_F(k)-n_F(k+\omega))$. Using the identity $(1+n_B(\omega))(n_F(k)-n_F(k+\omega))=n_F(k)(1-n_F(k+\omega))$, one can show the equivalence between Eq.~(\ref{interaction_rate}) and Eq.~(\ref{inrav2}). 

\begin{figure}[htbp]
\begin{center}
\includegraphics[width=0.49\linewidth]{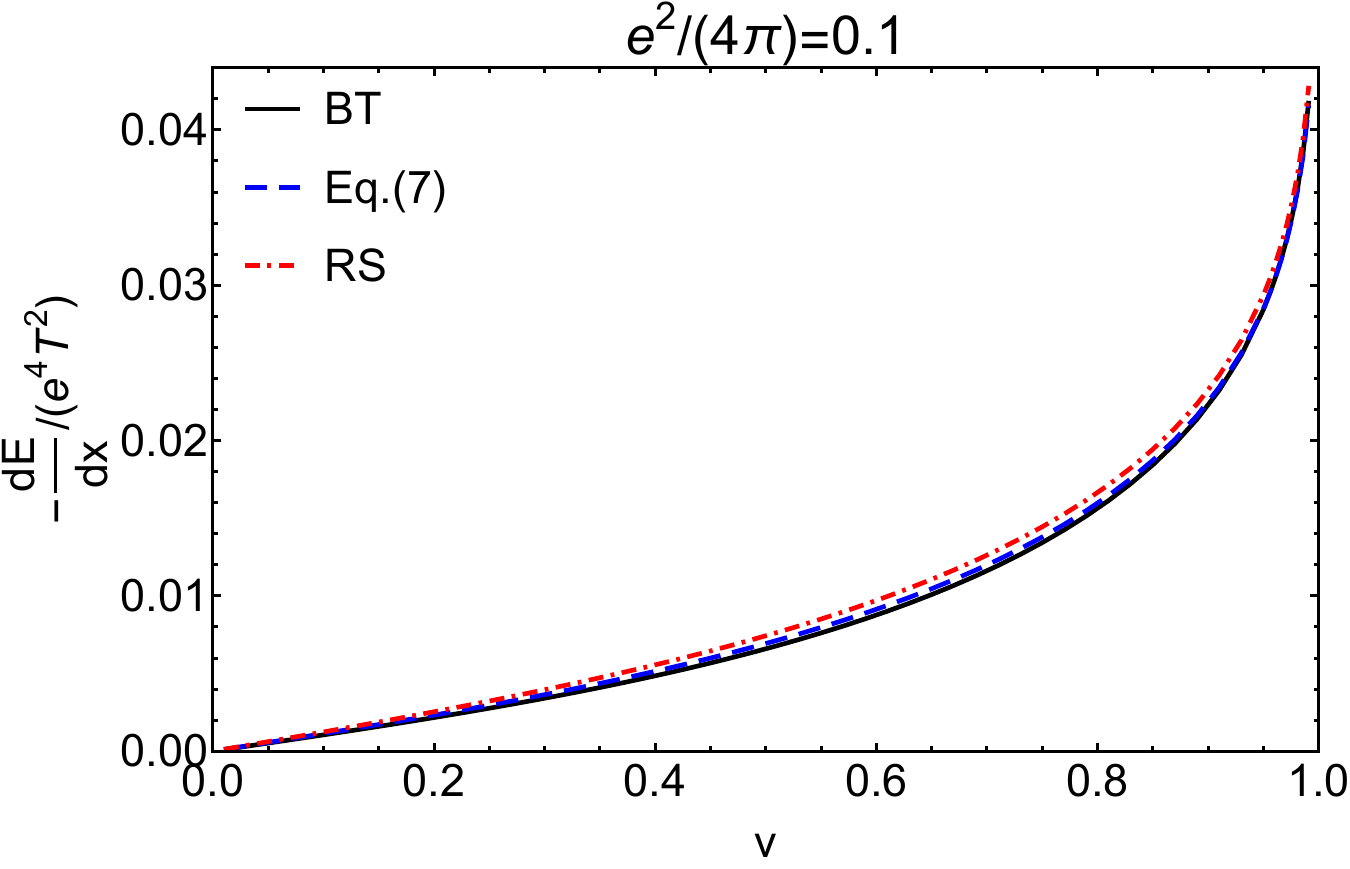}
\includegraphics[width=0.49\linewidth]{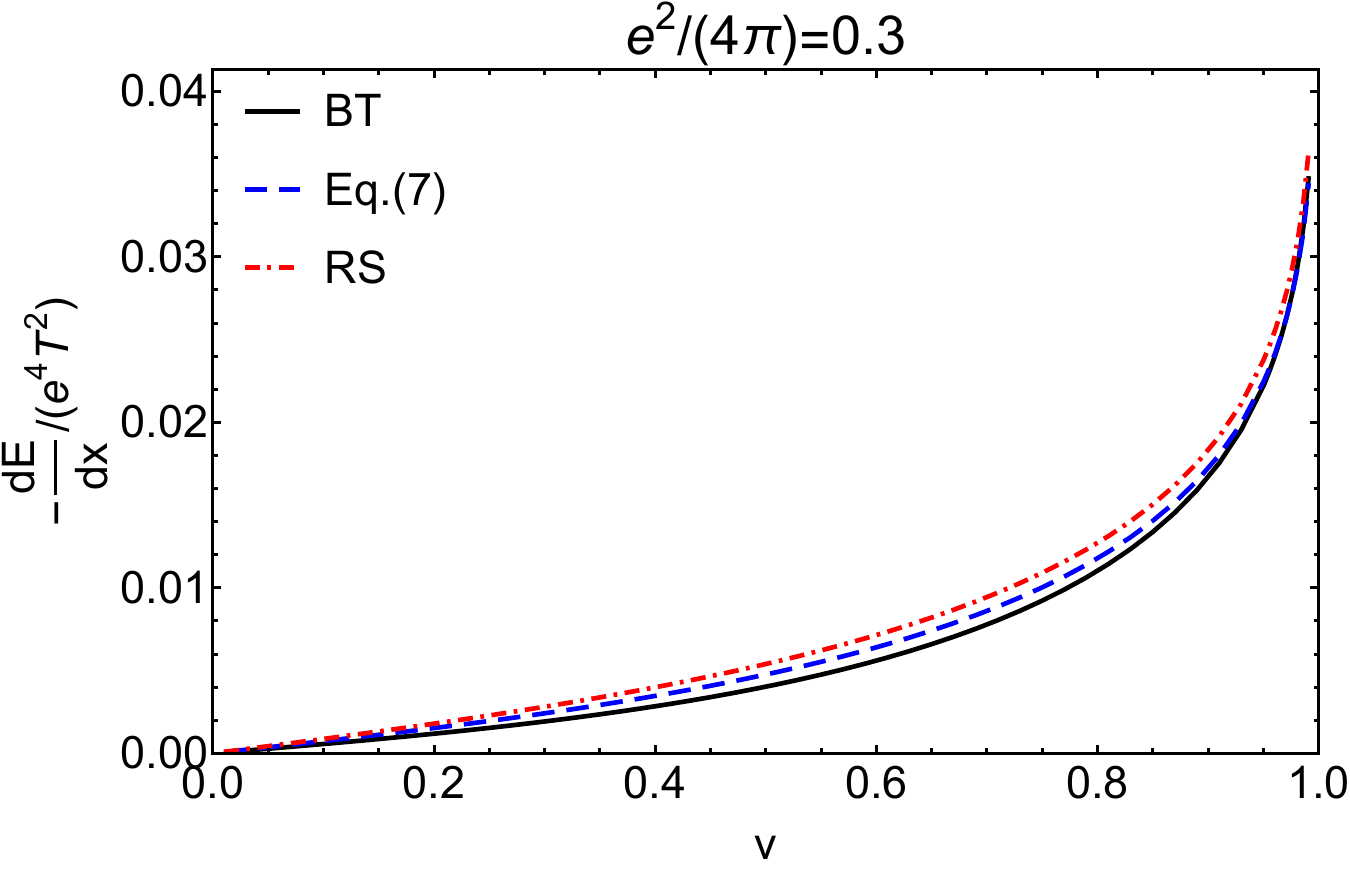}
\caption{Collisional energy loss as a function of $v$ for $e^2/(4\pi)=0.1$ (left) and $e^2/(4\pi)=0.3$ (right) obtained from different methods.}
\label{usdr2}
\end{center}
\end{figure}

According to the above discussions, we find that in the small coupling limit where $e T\ll q^\star \ll T$ can be well satisfied, there is good agreement for the total energy loss computed based on different theoretical methods. However, when extrapolating to moderate couplings, discrepancies appear. This has been numerically checked and the results are presented in Fig.~\ref{usdr2}. Notice that the RS result of the energy loss depends on the cutoff $q^\star$, we use the variational approach to eliminate this ambiguity. Therefore, the corresponding energy loss is the minimum of $-d E/ d x$ when varying $q^\star$.

Finally, we discuss the gauge dependence when using the HTL resummed propagator to calculate the squared matrix element. In general linear gauges, including covariant, Coulomb, and temporal axial gauges, the gauge-invariant part in the bare propagator is given by $-g^{\mu\nu}/Q^2$. On the other hand, the gauge-dependent terms are proportional to either $Q^\mu Q^\nu$ or $Q^\mu M^\nu+Q^\nu M^\mu$. It can be shown that these gauge dependent terms do not contribute to the squared matrix element. Notice that one Dirac trace associated with the incident heavy fermion is ${\rm Tr} [({\displaystyle{\not}P^\prime}+M)\gamma_\mu({\displaystyle{\not}P}+M)\gamma_{\mu^\prime}]$, when contracted with $Q^\mu$, one obtains zero. Similarly, the other Dirac trace associated with the medium light fermion is ${\rm Tr}[{\displaystyle{\not} K^\prime}\gamma_\nu {\displaystyle{\not} K} \gamma_{\nu^\prime}]$ which also leads to zero when contracted with $Q^\nu$. As a result, the squared matrix element is gauge-independent as expected.

The above analysis also applies to the case where the resummed propagator is used. Since it has the same Lorentz structure as the bare propagator, the gauge-dependent terms in the resummed propagator have no contributions for the same reason as discussed above. It should be pointed out that besides a term proportional to $g^{\mu\nu}$, there is another gauge-independent contribution proportional to $M^\mu M^\nu$ in the resummed propagator. Therefore, to calculate the matrix element, one only needs to consider the gauge independent terms in Eq.~(\ref{resumpro}) which are given by
\be\label{resumprogain}
{\tilde D}^{\mu\nu}(Q)=\frac{-g^{\mu\nu}}{Q^2-\Pi_T(\oh)} +\Bigg(\frac{1-\oh^2}{Q^2-\Pi_T(\oh)}+\frac{1}{q^2-\Pi_L(\oh)}\Bigg)M^\mu M^\nu \,. 
\ee

Using Eq.~(\ref{resumprogain}), we can calculate the squared matrix element. In the approximations that $M\gg T$ and $p\gg T$, the result can be expressed as\footnote{Notice that the squared matrix element obtained in Ref.~\cite{Braaten:1991we} has a wrong sign before the term $(k k^\prime-{\bf k} \cdot{\bf k}^\prime)$, but this sign is irrelevant to the soft contributions because the approximation ${\bf k}\approx {\bf k}^\prime$ holds.}
\ba\label{ms}
&&\frac{1}{2}\sum_{{\rm spin}}|{\cal M}|^2=16e^4\Big\{ |\Delta_L(Q)|^2 E^2 (k k^\prime +{\bf k} \cdot {\bf k}^\prime)\nonumber \\
&&\hspace{2cm}+2\,{\rm Re} (\Delta_L(Q) \Delta_T(Q)^*) E\big[k({\bf p} \cdot{\bf k}^\prime - ({\bf p} \cdot {\hat {\bf q}})( {\bf k}^\prime \cdot {\hat {\bf q}}) )+k^\prime ({\bf p} \cdot{\bf k} - ({\bf p} \cdot {\hat {\bf q}})( {\bf k}\cdot {\hat {\bf q}}) )\big]\nonumber \\
&&\hspace{2cm}+|\Delta_T(Q)|^2\big[2({\bf p} \cdot{\bf k} - ({\bf p} \cdot {\hat {\bf q}})( {\bf k} \cdot {\hat {\bf q}}) )({\bf p} \cdot{\bf k}^\prime - ({\bf p} \cdot {\hat {\bf q}})( {\bf k}^\prime \cdot {\hat {\bf q}}) )
\nonumber \\
&&\hspace{8cm}+(k k^\prime-{\bf k} \cdot{\bf k}^\prime)(p^2 - ({\bf p} \cdot {\hat {\bf q}})( {\bf p} \cdot {\hat {\bf q}}) )\big]\Big\}\,, 
\ea
where $ {\hat {\bf q}}={\bf q}/q$ and the second line does not contribute to the energy loss after averaging over the direction of ${\bf v}$.

\section{The collisional energy loss of a heavy fermion in a hot plasma with a BGK collisional kernel}\label{elwithcolli}

To incorporate the collision effect in the calculation of the fermion energy loss, a feasible way is to derive the photon self-energy from the kinetic equation with a specified collisional kernel, and then compute the corresponding resummed propagator through the Dyson-Schwinger equation. Based on Eq.~(\ref{elnew}), the collisionally-modified resummed propagator is apparently the key ingredient to study the energy loss in a collisional plasma. We use the BGK collisional kernel, which is given by
\be
{\cal C}({\bf k},X)=-\nu \Big[f({\bf k},X)-\frac{\int_{\bf k}f({\bf k},X)}{\int_{\bf k}n_{F}(k)}n_{F}(k)\Big]\, ,
\ee
where $\nu$ is the collision rate, which is inversely proportional to the equilibration rate of the plasma under collisions between the hard partons. The BGK collisional kernel ensures an instantaneously conserved number of particles, which improves the relaxation time approximation. In the above equation, we use the shorthand notation $\int_{\bf k}\equiv \int d^3{\bf k}/(2\pi)^3$ and $f({\bf k}, X)=f ({\bf{k}})+\delta f({\bf k}, X)$ where the fluctuation $\delta f({\bf k}, X)$ presents a slight deviation of the distribution function from its homogeneous values $f(\bf{k})$. 

According to the Maxwell equation, the induced current $J_{\rm ind}^\mu (X)$ is given by $J_{\rm ind}^\mu (X)= e \int_{{\bf k}} V^\mu \delta f ({\bf k}, X)$, and thus can be obtained by solving the linearized kinetic equation for the fluctuation $\delta f({\bf k}, X)$ which, in momentum space, reads
\be
i (- \omega + {\bf q} \cdot {\bf v}) \delta f({\bf k}, Q) \pm e V_\mu F^{\mu \nu} (Q) \partial_\nu f({\bf k}) ={\cal C} ({\bf k}, Q)\,  ,  
\ee
where $V=(1,{\bf v})$ with ${\bf v}={\bf k}/k$. The field strength tensor is $F^{\mu \nu}=\partial^\mu A^\nu-\partial^\nu A^\mu$. Furthermore, the $+$ and $-$ signs correspond to electrons and positrons, respectively.

The photon self-energy is determined by functional differentiation of the induced current with respect to the gauge field and the result can be expressed as
\ba\label{defpi}
\Pi^{\mu\nu} (Q)&=&\frac{\delta J_{{\rm ind}}^\mu(Q)}{\delta A_\nu(Q)}=e^2 \int_{{\bf k}} V^\mu \partial^{({\bf k})}_{l} f({\bf{k}})\frac{g^{l\nu}(\oh-\hat{{\bf{q}}}\cdot{\bf{v}})-{\hat Q}^l V^\nu}{\oh-\hat{{\bf{q}}}\cdot{\bf{v}}+i \uh}\, \nonumber \\
&+& e^2(i \uh) \int\frac{\mathrm{d} \Omega}{4\pi} \frac{V^\mu}{\oh-\hat{{\bf{q}}}\cdot{\bf{v}}+i \uh}\int_{{\bf k^\prime}} \partial^{({\bf k }^\prime)}_{l} f({\bf{k}}^\prime)\frac{g^{l\nu}(\oh-{\hat {\bf{q}}}\cdot{\bf{v}}^\prime)-{\hat Q}^l {V^\prime}^\nu}{\oh-{\hat {\bf{q}}}\cdot{\bf{v}^\prime}+i \uh} \mathcal{W}^{-1}(\hat \omega,\hat\nu ) \, .
\nonumber \\
\ea
In the above equation, ${\bf{v}}^\prime={\bf{k}}^\prime/k^\prime$. We also define the following dimensionless quantities $\hat\nu=\nu/q$ and ${\hat Q}=({\hat{\omega}, {\hat {\bf q}}})$. In addition, $\mathcal{W}(\hat \omega,\hat\nu )$ is given by
\be
\mathcal{W}(\hat \omega,\hat\nu )=1-\frac{i \uh}{2}\int_{-1}^{1}\mathrm{d} x\frac{1}{\oh-x+ i\uh}=1+\frac{i\uh}{2} \ln\frac{z-1}{z+1}\, .
\ee
with $z\equiv {\hat {\omega}} +i {\hat {\nu}}$.
As discussed in Ref.~\cite{Zhao:2023mrz}, it can be proven that the above self-energy is transverse so that $Q_\mu \Pi^{\mu\nu}=Q_\nu \Pi^{\mu\nu}=0$. However, it is not symmetric in Lorentz indices due to the appearance of the BGK collisional kernel. Notice that the distribution function $f(\bf{k})$ in Eq.~(\ref{defpi}) is completely arbitrary and an anisotropic hard parton distribution in
momentum space will lead to a rather complicated structure of the photon self-energy which requires five structure functions in the decomposition. 
However, such a symmetry can be restored if an isotropic distribution is considered. In this work, we consider the thermal equilibrium distribution, i.e., $f({\bf{k}})=2n_F(k)$ where the factor $2$ comes from taking into account electrons and positrons with vanishing chemical potential. As a result, the photon self-energy can be decomposed as $\Pi^{\mu\nu} (Q) = \Pi_T A^{\mu\nu} +\Pi_L {\hat{\omega}}^{2} B^{\mu\nu}$ where the transverse and longitudinal parts of the photon self-energy are given by
\be\label{pitl}
 \Pi_T ({\hat {\omega}},{\hat {\nu}})= \frac{m_\gamma^2}{4} {\hat{\omega}} \bigg[2z+(z^2-1)\ln\frac{z - 1}{z + 1}\bigg]\, ,\quad \Pi_L ({\hat {\omega}},{\hat {\nu}}) =-\frac{m_\gamma^2}{2} \frac{1 }{ \mathcal{W}(\hat \omega,\hat\nu )}\bigg(2+z {\ln\frac{z-1}{z+1}}\bigg) \,.
\ee

As compared to the collisionless limit, the Lorentz structure of the photon self-energy with the BGK collisional kernel is unchanged\footnote{The same is not true when an anisotropic distribution function $f({\bf k})$ is used.}, therefore, the corresponding transverse and longitudinal resummed propagators remain the same as those given in Eq.~(\ref{protl}), provided that one uses Eq.~(\ref{pitl}) for the photon self-energy. It is obvious that our previous discussions concerning the gauge invariance based on Eq.~(\ref{elnew}) still hold in the presence of the BGK collisional kernel. As a result, one can directly use Eq.~(\ref{elnew}) to evaluate the energy loss and investigate the influence of collisions between medium partons on $- dE/d x$.

In general, a QED plasma is weakly coupled because the typical values of the coupling constant are rather small. Consequently, the effects of collisions become negligible, and no significant influence on the energy loss can be expected. On the contrary, considering such an influence on the heavy-quark energy loss in a QCD plasma is certainly more interesting because the strong coupling constant $g$ could become moderate in a deconfined plasma with temperatures not far above the critical temperature. Similarly as the fermion energy loss, an energetic heavy quark may also lose energy when passing through the QGP by scattering off the light quarks and gluons. Up to a trivial color factor, the above results can be generalized to the quark-quark elastic scattering in QCD. We postpone the study on the quark-gluon scattering to the future, since it requires a new calculation of the squared matrix element as well as a nontrivial verification on the gauge invariance when the resummed propagator is used. 

In a collisional QCD plasma, we can estimate the heavy-quark energy loss due to quark-quark scattering based on Eq.~(\ref{elnew}) where the coupling $e$ should be replaced by $g$ and the gluon self-energy can be obtained from Eq.~(\ref{pitl}) with $m_\gamma^2$ set to be the two-flavor QCD screening mass $m_D^2=4 g^2 T^2/3$. In addition, a color factor $2/3$ should be also included. However, due to the lack of the contributions from the quark-gluon scattering, our results cannot serve as a quantitative assessment on $-d E/d x$. On the other hand, since we are interested in the collision effect on the energy loss of a heavy-quark, we can actually focus on the ratio of the energy loss with and without the collisions. It needs to be noted that the energy loss ratio based on quark-quark scattering can provide a qualitative estimate on the energy loss ratio for full QCD including both quark-quark and quark-gluon scatterings. This is because these two scattering processes have a roughly equal energy loss ratio. 

As a rough estimate, we can assume the momentum transfer is large enough for hard processes, while small enough for soft processes\footnote{In our calculation, there is no need to introduce a cutoff scale for the momentum transfer. However, one can formally define the hard and soft processes as we did in Fig.~\ref{comqs}.}. With only the hard contributions, the energy loss ratio should be very close to $1$ as the collision effect is negligible for large momentum transfer. Furthermore, when considering soft momentum transfers, the squared matrix element computed with the resummed gluon propagator becomes identical for both quark-quark and quark-gluon scatterings according to our previous discussions. As a result, with only the soft contributions, these two scattering processes also have an identical energy loss ratio because the collision-induced modifications are entirely encoded in $|\mathcal{M}|^2$ through the resummed gluon propagators $\Delta_L(Q)$ and $\Delta_T(Q)$. Combining the hard and soft contributions, a roughly equal energy loss ratio for both quark-quark and quark-gluon scatterings can be expected provided that in the collisionless limit, the relative importance of the hard and soft contributions to the energy loss has no significant difference between these two different scattering processes. This is found to be true according to the known results~\cite{Braaten:1991we}.

The effects of collisions are very sensitive to the value of the collision rate chosen in numerical evaluations. However, as a phenomenological model for equilibration, the BGK collisional kernel cannot be derived from first principles. Therefore, determining the collision rate $\nu$ seems to be a rather challenging task. On the other hand, because the energy loss has an explicit dependence on the coupling constant, introducing a $g$-dependent collision rate turns out to be very reasonable. In this work, we adopt the parametrization for the collision rate as used in previous literature~\cite{Schenke:2006xu}, \mbox{$\nu/T \approx 5.2 \alpha_s^2 \ln (1+ 0.25/\alpha_s)\,$} with $\alpha_s=g^2/(4\pi)$. Therefore, the dimensionless collision rate ${\tilde \nu}=\nu/m_D$ can be written as
\be\label{nu}
{\tilde \nu} \approx 1.27 \alpha_s^{3/2} \ln (c+ 0.25/\alpha_s)\, .
\ee
In the above equation, the constant $c$ varies from $1$ to $2$ in our numerical results. For a non-Abelian gauge theory which has the feature of asymptotic freedom, according to Eq.~(\ref{nu}), the collision rate vanishes in the high-temperature limit where $\alpha_s \rightarrow 0$. This is actually an expected behavior. In the temperature region close to the critical temperature, choosing a typical value of the coupling constant $\alpha_s =0.3$, the collision rate ${\tilde {\nu}} \approx 0.13 \sim 0.22$ which is also consistent with the values commonly used. We note that the collision rate \eqref{nu} is the one appropriate at the timescale associated with successive hard-momentum exchanges.  There are more frequent soft-momentum exchanges that result in small-angle scatterings and these are the source of the logarithm in Eq.~\eqref{nu}.  Here, the constant $c$ that adds to the logarithm is taken to be an adjustable parameter to gauge the sensitivity of our results to its precise value.

\begin{figure}[htbp]
\begin{center}
\includegraphics[width=0.49\linewidth]{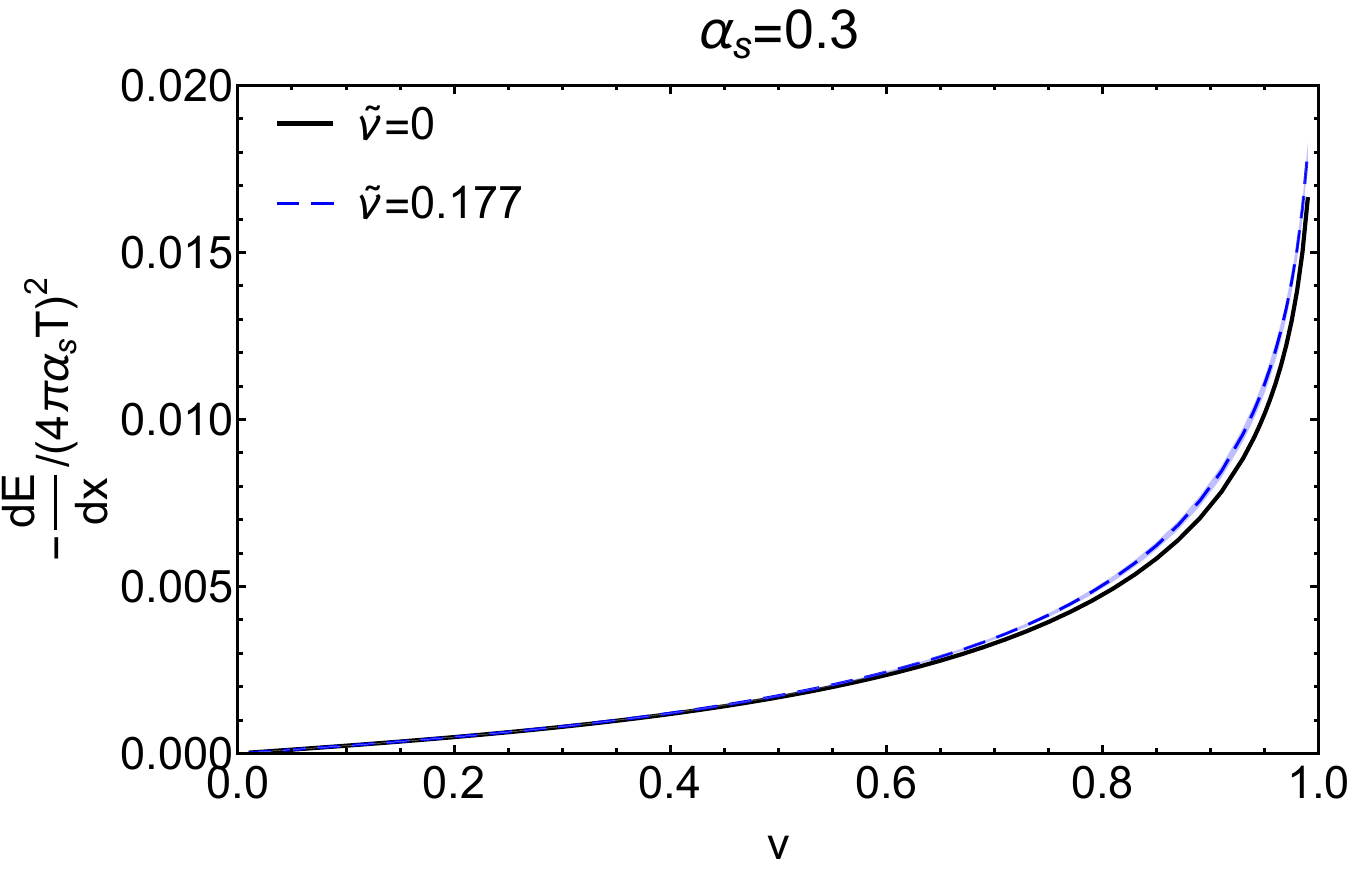}
\includegraphics[width=0.49\linewidth]{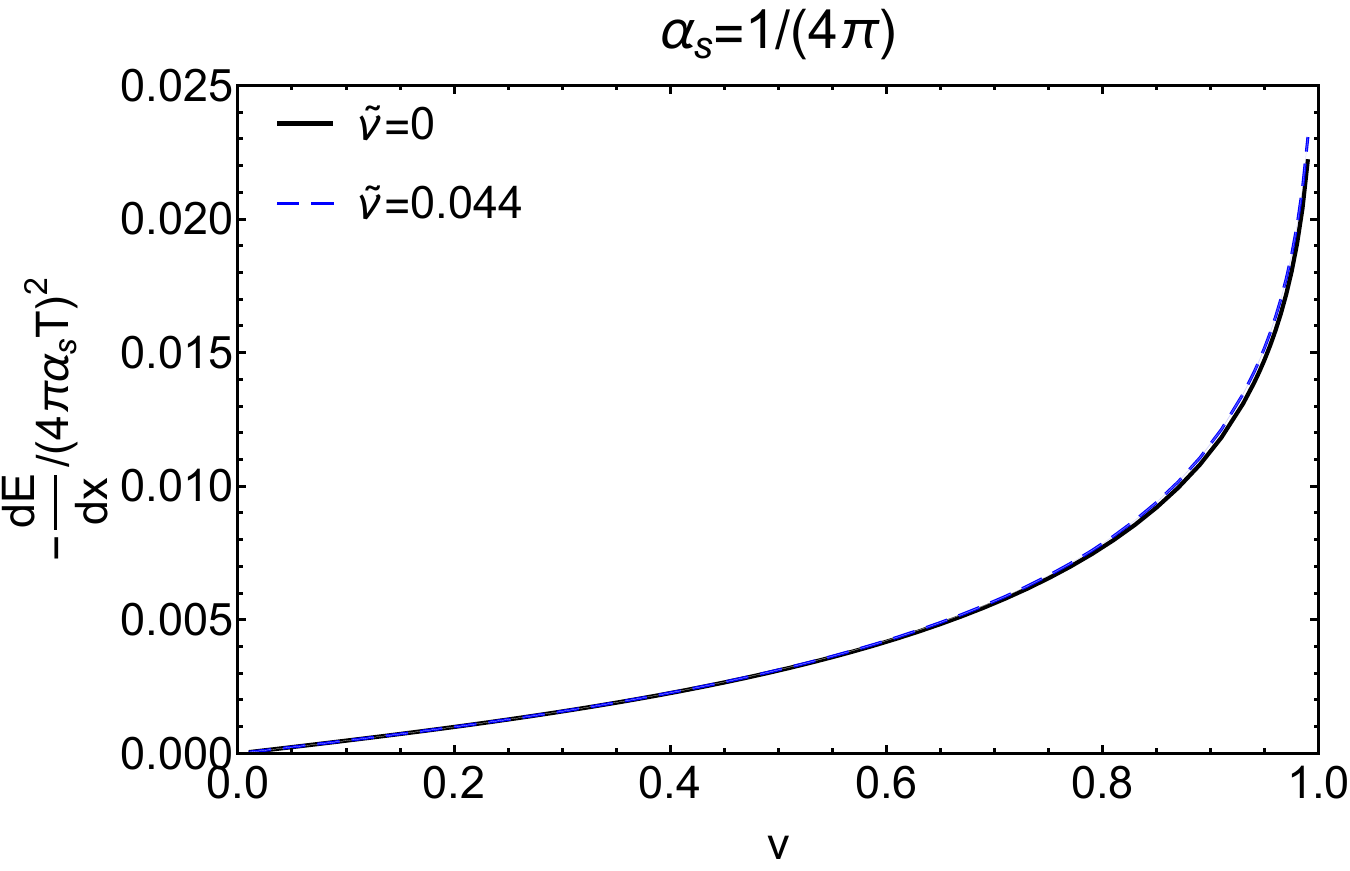}
\caption{Comparisons of the velocity dependence of the heavy-quark energy loss due to quark-quark scattering with and without collisions.  Left: $\alpha_s=0.3$ and the corresponding ${\tilde \nu}\approx 0.177$. Right: $\alpha_s=1/(4\pi)$ and the corresponding ${\tilde\nu}\approx 0.044$.}
\label{usdr}
\end{center}
\end{figure}

In Fig.~\ref{usdr}, we show the numerical results of the collisional energy loss due to quark-quark scattering as a function of the heavy-quark velocity. To avoid introducing any specified temperature dependence of the coupling constant which is not necessary for our purpose, we have scaled the energy loss by a factor $1/(4\pi\alpha_sT)^2$. For comparison, the scaled energy loss with and without the collisions are both presented in this figure. We choose two different values of the coupling constant, $\alpha_s =0.3$ and $\alpha_s=1/(4\pi)$. The former corresponds to a deconfined plasma near the critical temperature, where the collision rate is relatively large, ${\tilde \nu} \approx 0.177$ when taking $c=1.5$. The latter corresponds to the high-temperature limit, where the collision rate in the weakly coupled plasma becomes rather small, ${\tilde \nu}\approx 0.044$ with $c=1.5$. It is clear to see that given a small coupling constant applicable to very high temperatures, the collision rate determined by Eq.~(\ref{nu}) does not have a notable influence on the heavy-quark energy loss. However, for lower temperatures or larger collision rates, collisions among the medium partons can play a role in the evaluation of the collisional energy loss. Despite the lack of the explicit calculation of the quark-gluon scattering, we can still expect a qualitatively similar behavior between the two different scattering processes. Therefore, in general the heavy quark loses more energy in a collisional plasma and the enhancement of $-d E/d x$ is more significant when the incident velocity $v$ becomes large.

\begin{figure}[htbp]
\begin{center}
\includegraphics[width=0.49\linewidth]{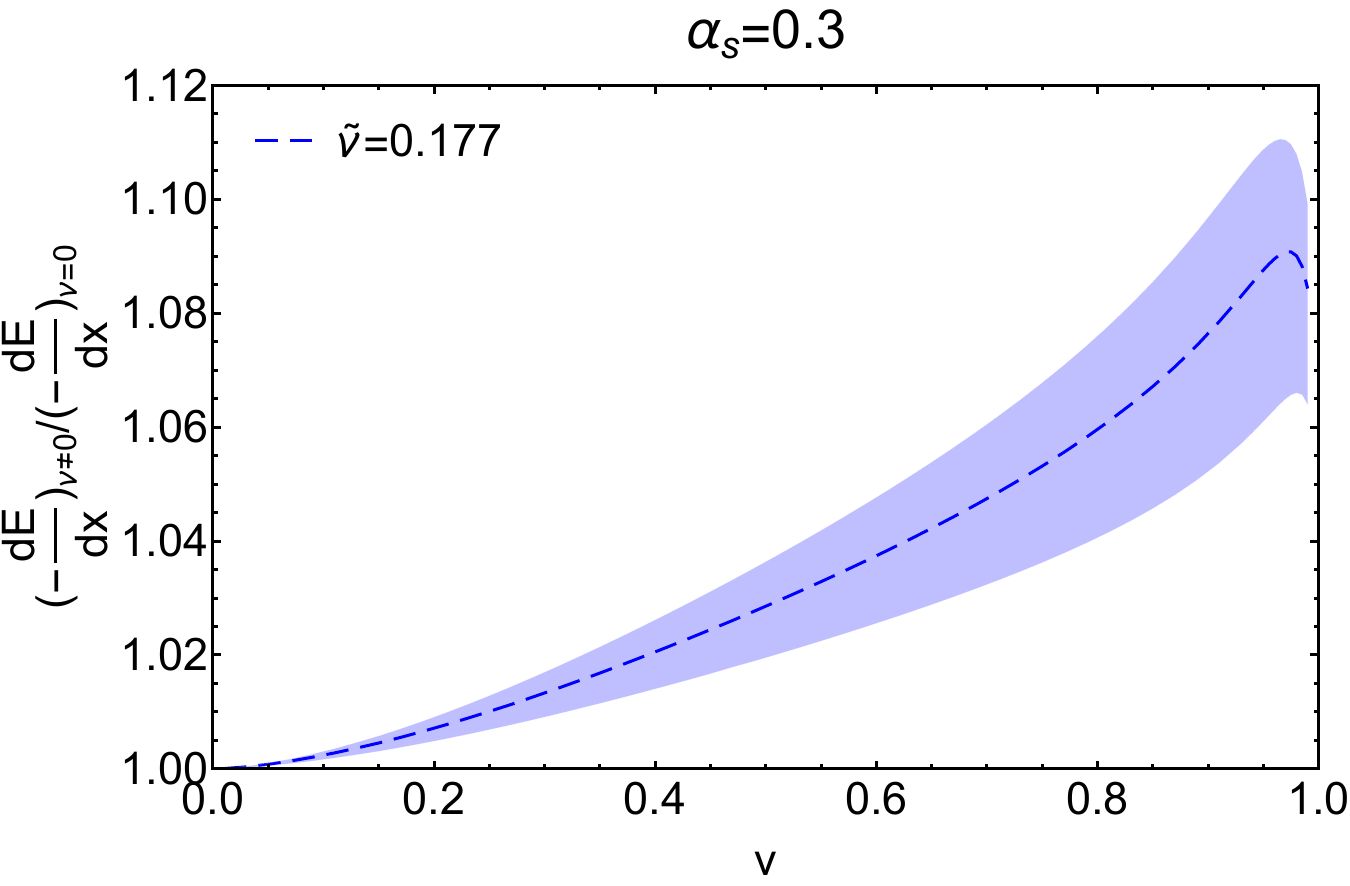}
\includegraphics[width=0.49\linewidth]{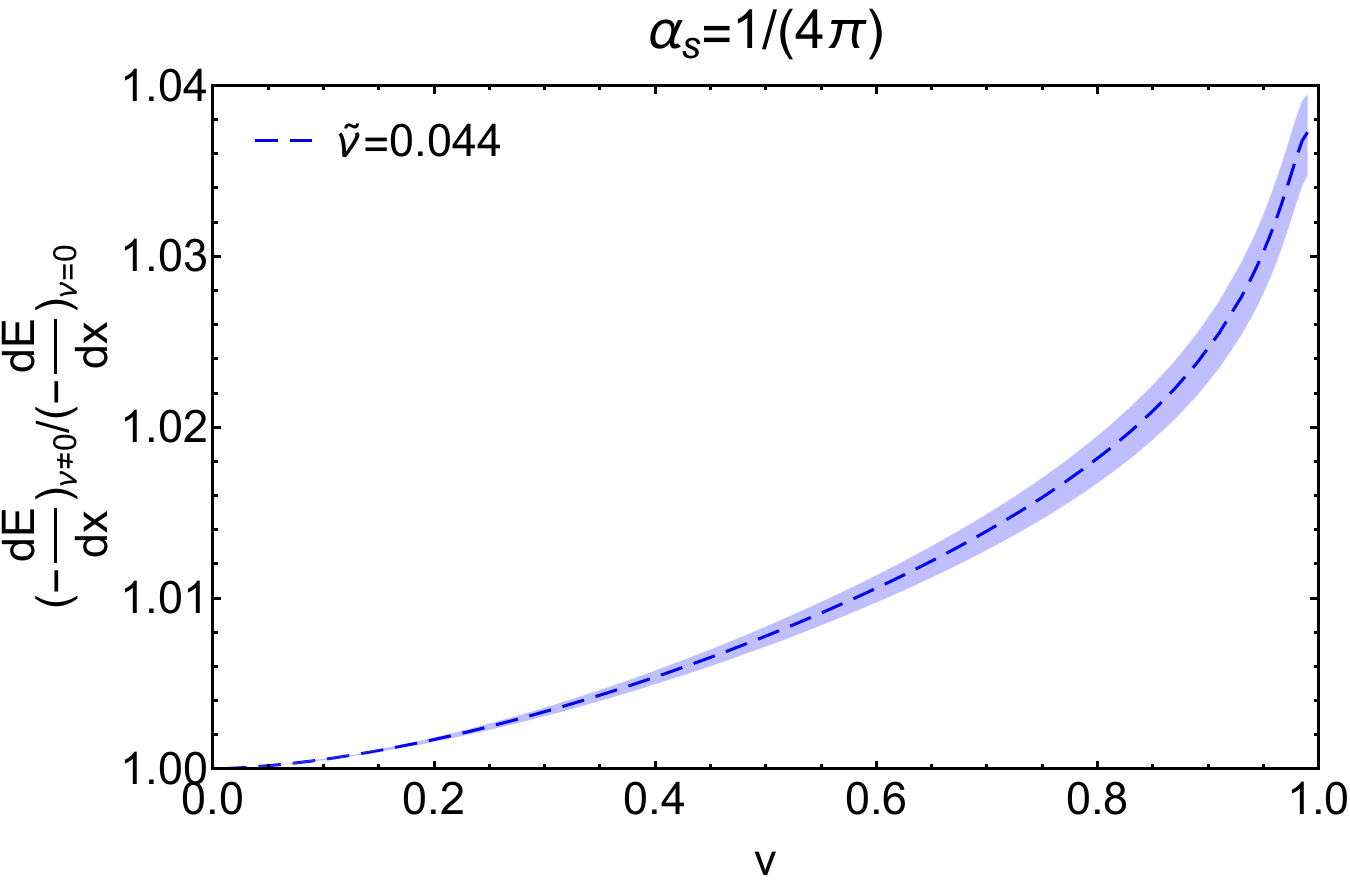}
\caption{The energy loss ratio as a function of the heavy-quark velocity. Left: $\alpha_s=0.3$ and the corresponding ${\tilde\nu}\approx 0.177$. Right: $\alpha_s=1/(4\pi)$ and the corresponding ${\tilde\nu}\approx 0.044$.}
\label{ravsv}
\end{center}
\end{figure}

As we already mentioned before, a more direct way to see the effects of collisions on the heavy-quark energy loss is to study the ratio of $-dE/dx$ with and without collisions. Our numerical results in Fig.~\ref{ravsv} further confirm that in the high-temperature limit, the collisions among thermal partons result in corrections to $-d E/d x$ are at most $\sim 5 \%$ when $\alpha_s=1/(4\pi)$. On the other hand, for a realistic coupling constant near the critical temperature, the influence on the energy loss becomes moderate, and the corrections can reach $\sim 10 \%$ for large incident velocities. Note that these magnitudes are strongly dependent on the parameterization of the collision rate $\nu$ as we used in the evaluation. Therefore, a more accurate determination of the values of $\nu$ turns out to be very crucial to draw a quantitative conclusion on the energy loss in a collisional plasma. In addition, we also find that although it has a significant deviation from unity only in the large $v$ region, the energy loss ratio is not a monotonic function of the velocity and there exists a turning point at some very large velocity where the effects of collisions are most pronounced. This is actually true for both small and large coupling constants.

\begin{figure}[htbp]
\begin{center}
\includegraphics[width=0.49\linewidth]{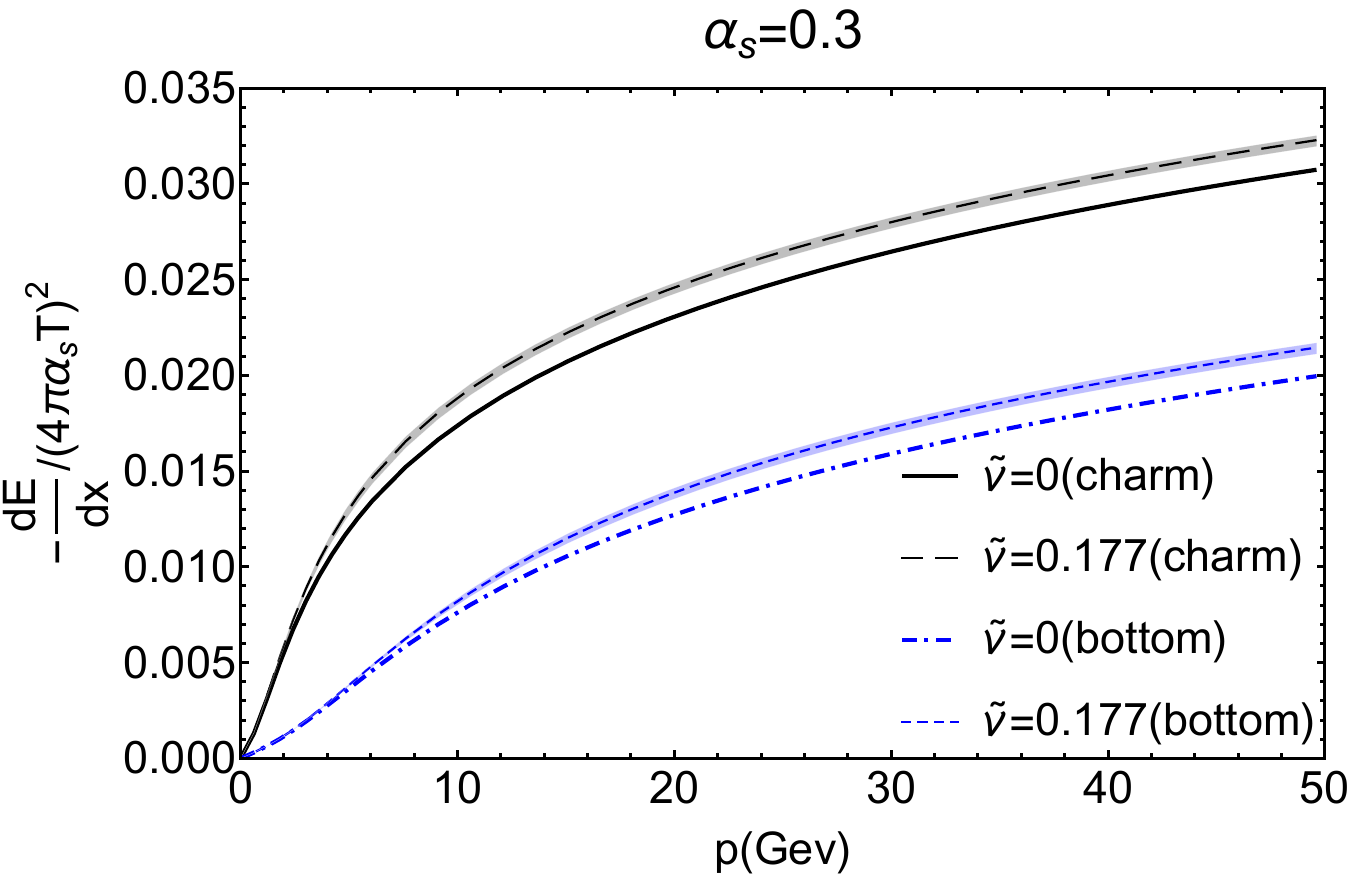}
\includegraphics[width=0.49\linewidth]{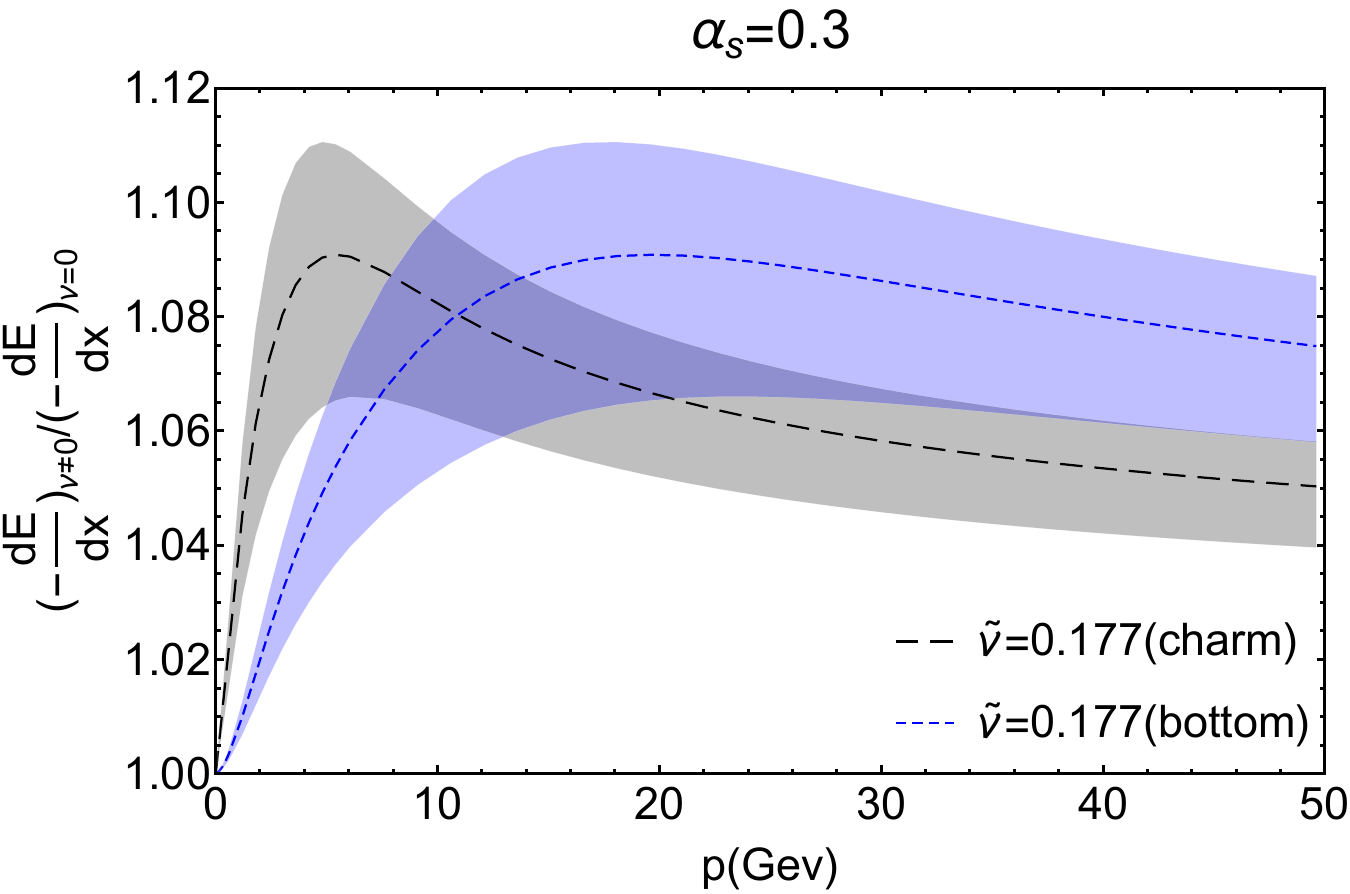}
\caption{Left: comparisons of the momentum dependence of the energy loss due to quark-quark scattering with and without collisions for charm and bottom quark at $\alpha_s=0.3$.  Right: the energy loss ratio as a function of the momentum for charm and bottom quark at $\alpha_s=0.3$.}
\label{elcandb}
\end{center}
\end{figure}

In order to see the flavor dependence of the collision effects on the heavy-quark energy loss, in Fig.~\ref{elcandb} we show the scaled energy loss due to quark-quark scattering and the energy loss ratio as a function of the heavy-quark momentum $p$ for both charm and bottom quarks with a strong coupling constant $\alpha_s=0.3$. The quark masses are chosen to be $m_c =1.3\,{\rm GeV}$ and $m_b =4.7\,{\rm GeV}$ for the charm and bottom quark, respectively. In general, the mass hierarchy of the collisional energy loss, i.e., that the charm quark loses more energy than the bottom quark, is also observed in a collisional plasma. Roughly speaking, when compared to the energy loss in a collisionless plasma, the increase in $-dE/dx$ of a charm quark is comparable to that of a bottom quark when the momentum becomes large. Therefore, a more significant correction to the energy loss can be expected for the bottom quark in the large-momentum region. However, in the small $p$ region, the opposite is true according to the plot on the right-hand side of this figure. Of course, the maximum of the energy loss ratio has no flavor dependence, a $\sim 10 \%$ correction induced by the collisions among thermal partons is found for both charm and bottom quark. 

Finally, we compare our results with those obtained in Ref.~\cite{Han:2017nfz} where the authors also considered the heavy-quark energy loss in the quark-gluon plasma with the same BGK collisional kernel. However, with a similar collision rate $\nu \approx 0.1\sim 0.2 m_D$, the corrections to $-dE/dx$ in the collisionless limit found in \cite{Han:2017nfz} are much larger than $\sim 10 \%$. The discrepancy originates from the different theoretical frameworks adopted in these two works. In Ref.~\cite{Han:2017nfz}, the energy loss formula given by Eq.~(\ref{soft}) was used to calculate $-d E/d x$, and thus one has to introduce an upper cut for the transferred momentum to eliminate logarithmic divergence in the final results. According to the discussions in Sec.~\ref{re}, this is because this formula cannot properly deal with the hard processes. The maximum of the transferred momentum was simply set to be the Debye mass in \cite{Han:2017nfz}, therefore, we can naturally conjecture that the overestimated corrections to the energy loss are related to the missing contributions from hard processes. Unlike the soft contributions, self-energy insertion into the bare propagator, which encodes all the information about the collisions among thermal partons, becomes less accentuated for scatterings with large momentum transfer. In the current work, we treat both the hard and soft scatterings in a unified framework, which self-consistently includes the effect of collisions and is free of any artificial cutoff. As a result, a more reliable estimate of the corrections to the heavy-quark energy loss can be expected.

\section{Conclusions and Outlook}\label{con}

In this work, we considered the collisional energy loss of a high-energy fermion passing through a hot and dense plasma. The equilibration of the plasma was described by the BGK collisional kernel, which led to a modification on the photon/gluon collective modes and thus affected the propagator in the resummed perturbation theory. In particular, we studied the effects of collisions on the energy loss of a fast fermion in a QED plasma by calculating the contributions from both hard and soft scatterings in a unified theoretical framework where collision effects were self-consistently encoded in a modified hard-thermal-loop resummed propagator. Based on our results, we investigated the heavy-quark energy loss in a collisional quark-gluon plasma through a simple generalization from QED to QCD. Numerical results showed that the energy loss of a heavy quark increased after including collisions among medium partons. The magnitude of the increase became negligible in the weak-coupling limit. However, near the critical temperature, according to our parametrization of the collision rate, taking into account the collisions gave rise to a moderate correction to the heavy-quark energy loss in the collisionless limit, which could result in a $\sim 10 \%$ increase at large incident velocities. In addition, for heavy quarks carrying large momenta, the collision-induced correction was more pronounced for a bottom quark, while the opposite occurred in the small momentum region where the energy loss of a charm quark became more sensitive to the collisions. Irrespectively of this, a mass hierarchy of the energy loss was observed in a collisional plasma.
 
Although the method adopted in this work did not require the introduction of an explicit separation scale for the momentum exchanges, there are alternative theoretical approaches to computing the collisional energy loss including the effects of both hard and soft exchanges. We made systematic comparisons with these different approaches and demonstrated that, in the weak-coupling limit, the results for the collisional energy loss obtained from various theoretical approaches became identical. In this limit, the dependence on a cutoff $q^\star$ introduced in Ref.~\cite{Braaten:1991we}, which separates the hard and soft processes, cancels exactly when considering the total energy loss. However, when the coupling constant was increased, we found differences among these approaches, and an uncertainty in the energy loss related to the choice of the separation scale also emerged. In addition to not requiring the introduction of a separation scale, when compared with other approaches, one important advantage of the approach used herein was that the self-energy insertion naturally goes to zero in the high-momentum limit. Therefore, there was no sharp transition from the soft processes involving a self-energy resummation to the hard processes, where instead the bare propagator was used.  Finally, we presented a proof that our results are manifestly gauge-invariant.
 
A complete calculation of the heavy-quark energy loss in a collisional QCD plasma still needs to be carried out in the future where the gauge invariance of contributions from quark-gluon scatterings should be considered when using a resummed gluon propagator. In addition, determination of the collision rate of the quark-gluon plasma in a concrete manner is very important, not only for the evaluation on the energy loss, but also for many other phenomenological studies.

In closing, we note that the coupling dependence of the collision rate found herein suggests that a moderate enhancement of the energy loss is expected only for temperatures not far above the critical temperature, which is termed as ``semi"-QGP due to the fact that the QGP may only be partially deconfined at these temperatures~\cite{Hidaka:2008dr}. We point out that in such a partially deconfined phase, nontrivial holonomy for Polyakov loops could also affect the collisional energy loss significantly~\cite{Lin:2013efa,Du:2024riq}. Therefore, a comprehensive understanding of this issue will be challenging, and further work will be needed in the future.

\section*{Acknowledgments}
The work of Y.G. is supported by the NSFC of China under Project No. 12065004 and by the Central Government
Guidance Funds for Local Scientific and Technological
Development, China (No. Guike ZY22096024). M.S. was supported by the U.S. Department of Energy, Office of Science, Office of Nuclear Physics Award No.~DE-SC0013470.

\bibliographystyle{apsrev4-1}
\bibliography{paper}

\end{document}